\documentclass{aa} 
\usepackage{graphicx}
\usepackage{txfonts}
\usepackage{url}
\usepackage{amsmath}
\usepackage{natbib,twoopt}
\usepackage[breaklinks=true]{hyperref} %% to avoid \citeads line fills
\hypersetup{colorlinks=true,urlcolor=blue,citecolor=blue,pdfborder= 0 0 0}
\bibpunct{(}{)}{;}{a}{}{,}  %% natbib format for A&A and ApJ
\makeatletter
 \newcommandtwoopt{\citeads}[3][][]{\href{http://adsabs.harvard.edu/abs/#3}%
 {\def\hyper@linkstart##1##2{}%
 \let\hyper@linkend\@empty\citealp[#1][#2]{#3}}}
 \newcommandtwoopt{\citepads}[3][][]{\href{http://adsabs.harvard.edu/abs/#3}%
 {\def\hyper@linkstart##1##2{}%
 \let\hyper@linkend\@empty\citep[#1][#2]{#3}}}
 \newcommandtwoopt{\citetads}[3][][]{\href{http://adsabs.harvard.edu/abs/#3}%
 {\def\hyper@linkstart##1##2{}%
 \let\hyper@linkend\@empty\citet[#1][#2]{#3}}}
 \newcommandtwoopt{\citeyearads}[3][][]%
 {\href{http://adsabs.harvard.edu/abs/#3}
 {\def\hyper@linkstart##1##2{}%
 \let\hyper@linkend\@empty\citeyear[#1][#2]{#3}}}
\makeatother

\begin{document}

 \title{Properties of brightest group galaxies in cosmic web filaments}

 \author{Teet Kuutma\inst{1}
     \and Anup Poudel\inst{2}
     \and Maret Einasto\inst{1}
     \and Pekka Hein\"am\"aki\inst{2}
     \and Heidi Lietzen\inst{1}
     \and Antti Tamm\inst{1}
     \and Elmo Tempel\inst{1}
%     \and others\inst{1}.
         %\fnmsep\thanks{Just to show the usage of the elements in the author field}
  }

 \institute{Tartu Observatory University of Tartu, Observatooriumi~1, 61602 T\~oravere, Estonia\\
  \email{teet.kuutma@to.ee}
  \and
  Tuorla Observatory, Department of Physics and Astronomy, Vesilinnantie 5, University of Turku, 20014 Turku, Finland
 }

 \date{}

% \abstract{}{}{}{}{} 
% 5 {} token are mandatory
 
 \abstract
 {The cosmic web, a complex network of galaxy groups and clusters connected by filaments, is a dynamical environment in which galaxies form and evolve. However, the impact of cosmic filaments on the properties of galaxies is difficult to study because of the much more influential local (galaxy-group scale) environment.
 }
 {The aim of this paper is to investigate the dependence of intrinsic galaxy properties on distance to the nearest cosmic web filament, using a sample of galaxies for which the local environment is easily assessable.}
 {Our study is based on a volume-limited galaxy sample with $M_\mathrm{r}$ $\leq -19$ mag, drawn from the SDSS DR12. We chose brightest group galaxies (BGGs) in groups with two to six members as our probes of the impact of filamentary environment because their local environment can be determined more accurately. We use the Bisous marked point process method to detect cosmic-web filaments with radii of $0.5-1.0$ Mpc and measure the perpendicular filament spine distance ($D_{\mathrm{fil}}$) for the BGGs. We limit our study to $D_{\mathrm{fil}}$ values up to 4 Mpc. We use the luminosity density field as a tracer of the local environment. To achieve uniformity of the sample and to reduce potential biases we only consider filaments longer than 5 Mpc. Our final sample contains 1427 BGGs.
 }
 {We note slight deviations between the galaxy populations inside and outside the filament radius in terms of stellar mass, colour, the 4000~\AA~break, specific star formation rates, and morphologies. However, all these differences remain below 95\,\% confidence and are negligible compared to the effects arising from local environment density.
 }
 {Within a 4 Mpc radius of the filament axes, the effect of filaments on BGGs is marginal. The local environment is the main factor in determining BGG properties.
 }
 
 \keywords{galaxies: general -- galaxies: statistics -- galaxies: evolution -- galaxies: groups: general -- large-scale structure of Universe}

 \maketitle
%
%________________________________________________________________
\section{Introduction}
On large scales the distribution of galaxies forms a complex network of structures composed of dense knots, elongated filaments, walls, and underdense voids. This network, called the cosmic web, formed as a result of perturbations in the matter distribution of the early Universe \citep[e.g.][]{Zeldovich:70, Bond:96, vandeWeygaert:08}. The most visually striking features of this web structure are the dense knots, which are the sites of galaxy clusters, and the filaments, elongated chains of galaxies connecting the knots. However, delineating filaments on the basis of the distribution of galaxies is complicated. Observable galaxies and their dark matter halos represent only about half of the total matter contained in filaments \citep{Haider:16}. Furthermore, observational techniques cannot provide the full 6D phase space information needed for an accurate reconstruction of the cosmic web. Nevertheless, several computational methods have been proposed for extracting cosmic web structures; see \citet{Libeskind:18} for a recent overview and comparison of simulations, and \citet{Rost:20} for a comparison of observations.

Cosmological dark matter simulations \citep[e.g.][]{Gramann:93, Springel:05a, Cautun:14} have shown that the cosmic web forms as matter collapses anisotropically into increasingly dense and symmetric structures. Walls, filaments, and knots represent regions that have gravitationally contracted in one, two, and three dimensions, respectively, while void regions have been too sparse for 
% that. 
this contraction. 
Out of these four large-scale structure element types, the largest fraction of dark and baryonic matter is contained within filaments \citep{Aragon-Calvo:10}. Therefore, to understand galaxy formation and evolution, it is important to ask to what extent is the filament environment responsible for influencing the properties of galaxies it contains?

Filaments are known to have a statistical effect on the alignment of galaxies. Observational studies \citep{Jones:10, Tempel:13a, Tempel:13b, Hirv:16, Pahwa:16, Rong:16} have shown that the spin axis of spirals and the longer axis of elliptical galaxies tend to align with filaments. Beyond the galaxy scale, alignment has also been found between filaments and galaxy pairs \citep{Tempel:15b} and larger satellite systems \citep{Tempel:15a}. Two effects can cause galaxy alignments: gas inflow to galaxies at a preferred angle \citep{Tempel:14c, Libeskind:15} and tidal torques during the gravitational collapse \citep{Ganeshaiah_Veena:19}. 

Apart from the alignment, the dependence of other properties of galaxies on the filament environment is difficult to disentangle from the much stronger effects of the local (galaxy-group scale) environment. For example, the well-known morphology--density relation implies that galaxies in denser environments are preferentially of earlier types 
\citep{Hubble:31, Zwicky:38, Dressler:80, Postman:84},
% \citep{Hubble:31, Zwicky:38, Dressler:80},
are more massive, and are more passive \citep{Kauffmann:04, Gomez:03}. This relation is a result of the combination of several effects, which can be broadly split into the effects related to the galaxy mass and initial birthplace on the one hand, and to the cluster environment on the other, the latter being the most important for galaxies entering the cluster at later stages \citep[e.g.][]{Gunn:72, Moore:96, Peng:10}. While filaments generally represent a denser environment compared to walls and voids, but less dense than clusters, we can expect an impact of the environment density to be present. However, filaments can be found over a wide range of large-scale densities, from void environments to superclusters \citep[][]{Aragon-Calvo:10, Sousbie:11, Einasto:12, Alpaslan:14, Cautun:14, Darvish:16, Malavasi:17}, complicating a straightforward assessment of the effects arising from density alone. Another complicating effect is `galaxy conformity', an observation showing that galaxy properties are strongly dependent on those of their nearest massive neighbours \citep[e.g.][]{Weinmann:06,Park:07,Hearin:16}.

Despite the given difficulties, several recent studies have reached the conclusion that filament galaxies tend to be more massive, redder, and less star forming than galaxies outside filaments, suggesting that in addition to the effects on alignment, cosmic matter flows and tidal torques related to filaments can also alter the intrinsic properties of galaxies. These effects have been measured based on distance to the nearest filament axis at $z < 0.25$, \citep[e.g.][]{Alpaslan:16, Kuutma:17, Kraljic:18} and also at redshifts up to $z < 0.9$, \citep[e.g.][]{Chen:17, Malavasi:17, Laigle:18, Sarron:19}. Filament-based quenching in galaxies has also been observed  in studies that look for the impact of the large-scale structure, but are not concentrated specifically on filaments \citep[e.g.][]{Alam:19, Salerno:19}. Additionally, \citet{Odekon:18} found that HI content of galaxies increases with filament spine distance. They also found that the most gas rich galaxies at fixed local density and stellar mass are located in smaller `tendril' filaments inside voids. The interesting discovery of H$\alpha$-emitting gas clouds in the remote outskirts of filament galaxies \citet{Vulcani:19} also indicate the possibility of filament-induced gas-related transformations.

Implications have also been made  from theoretical considerations of filament formation. \citet{Musso:18} studied analytically how the filament environment affects halo masses, accretion rates, and formation times, finding that halo properties change significantly along the perpendicular direction from the filament spine: halo masses increase by two orders of magnitude, while mass accretion and formation times are 20--30\,\% higher and lower, respectively, for halo masses of $10^{11}~h^{-1}~M_\odot$. \citet{Kraljic:19} used the Horizon-AGN simulation to show how angular momentum and feedback processes shape velocity dispersions and specific star formation rates (SSFRs) of filament galaxies beyond the expected mass and density effects. In addition to perpendicular distance from filament spines, both works also consider distance from the filament saddle point along the filament axis. Qualitatively, their results agree with observations of stellar mass increase and SSFR decrease towards filament spines.

In contrast, some studies reach the conclusion that galaxy and halo properties depend solely on local density and that filament environment has no additional effects beyond the ones related to the local density enhancement. \citet{Yan:13} found the density (not the  anisotropy) of the potential field of galaxies to be the only determining factor of galaxy properties. \citet{Eardley:15} found that changes in galaxy luminosity functions in different large-scale environments depended solely on the local density. \citet{Brouwer:16} measured halo masses using galaxy-galaxy lensing and found no dependence of the average halo mass of central galaxies on their cosmic web environment. \citet{Goh:19} found in simulation data that at a given environmental density the cosmic web does not affect the shape or mass accretion of dark matter halos. These results imply a controversy in the understanding of the large-scale effects on galaxy evolution.

Few of the above-mentioned studies look at specific sub-populations of galaxies. However the effects of filaments on brightest group galaxies (BGGs) deserve separate consideration, because BGGs represent a distinct population of galaxies from the other group galaxies. Previous studies have shown that because of their exceptional brightness BGGs are reliable tracers of complete samples of groups. According to simulations BGGs are typically located in or near the centres of group potential wells (\citealt{Lange:18} and references therein). \citet{Shen:14} determined that on average BGGs are 20\% more massive than other group members.

Earlier studies about BGGs and their large-scale environment have found that the supercluster environment influences BGGs. Studies have covered individual superclusters and detailed studies of BGG populations in them \citep[e.g.][]{Einasto:11} and also supercluster catalogues \citep[e.g.][]{VonDerLinden:07}. The complicated merger, accretion, and feedback mechanisms in the denser supercluster regions leave an imprint on the evolution of BGGs in these environments \citep{Luparello:15, Li:19}. Looking at studies of direct connections between BGGs and filaments \citet{Poudel:17} found that BGGs in filaments are generally more massive and redder than those outside filaments at fixed group mass and large-scale (supercluster-scale) environments. They selected galaxy groups with at least five members in order to have acceptable dynamical mass estimates for groups.

In recent years the method of filament connectivity has been utilised to define the connection between galaxy groups and filaments \citep{DarraghFord:19, Kraljic:20}. It has been determined that there is a correlation with the number of filaments connected to a galaxy group and some group properties. Namely groups with more connectivity are on average more massive and contain less active galaxies. \citet{DarraghFord:19} have found that BGG mass correlates with connectivity in low-mass groups while for high-mass groups there is anti-correlation. 

We take a closer look at galaxies in and near filaments in an attempt to further quantify possible effects of filament environment on galaxy properties. Our aim is to check for filament effects after accounting for the local environment of galaxies. For filament detection we use the Bisous point-process method, and define filament distance as the perpendicular distance of a galaxy from the nearest filament spine ($D_{\mathrm{fil}}$). We use galaxies selected from the Sloan Digital Sky Survey (SDSS) Data Release 12 (DR12). We limit our study to intermediate environment densities, where the filaments found with the Bisous method are most reliable. We use the luminosity density field at BGG locations as a measure of the local environment, additionally limiting our study to BGGs. Unlike \citet{Poudel:17}, in this paper we focus on BGGs of lower mass groups. We look for differences in star formation tracers, colours, and morphologies as a function of $D_{\mathrm{fil}}$. The data used is described in Sect. \ref{sect:data}. In Sect. \ref{sect:meth} we give an overview of the Bisous filament finder method, describe how the environment of galaxies is determined, and discuss further considerations made when selecting BGGs for the analysis. Our results are given in Sect. \ref{sect:res}, and we conclude in Sect. \ref{sect:disc}. Throughout the paper our results are shown assuming Planck $\Lambda$CDM cosmology \citep{Planck:16}: $H_0 = 67.8~\mathrm{km~s^{-1}~Mpc^{-1}}$, $\Omega_\mathrm{m}$ = 0.308, and $\Omega_{\Lambda}$ = 0.692~\footnote{When referencing numerical values from other works that use $H_0 = 100~h~\mathrm{km~s^{-1}~Mpc^{-1}}$, the values are given with the corresponding $h$ parameter along with a conversion to the Planck cosmology used in this work.}.

\begin{figure}
 \centering
 \includegraphics[width=88mm]{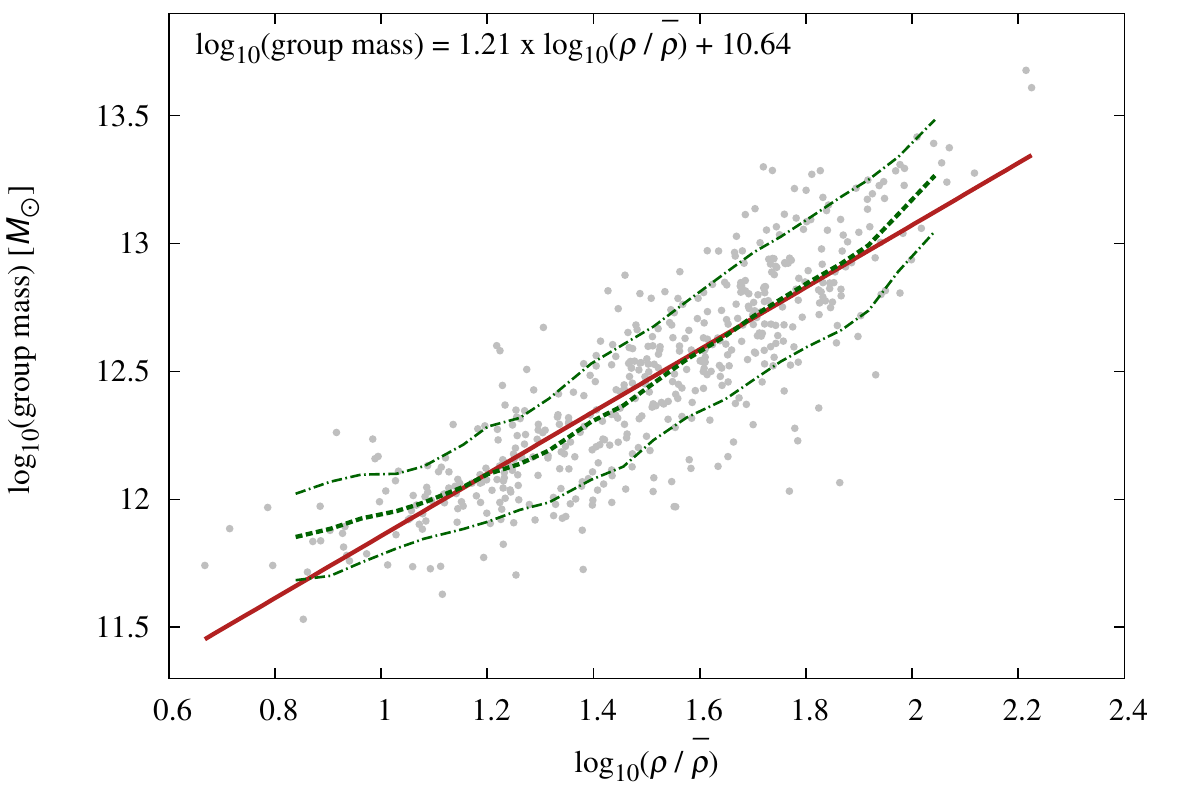}
 \caption{Group mass vs luminosity density of selected BGGs (grey points) in the EAGLE simulation. The thick green dashed line shows the median of a Gaussian fit to the group mass of the data points at different density values, while the thinner green dash-dotted lines show $1\sigma$ standard deviations from the median. The relation between the logarithms of group mass and luminosity densities shows a clear correlation that can be approximated by a linear fit, shown with the red line and the corresponding formula in the top of the figure.}
 \label{fig:eag_masslum}
\end{figure}

\begin{figure}
 \centering
 \includegraphics[width=88mm]{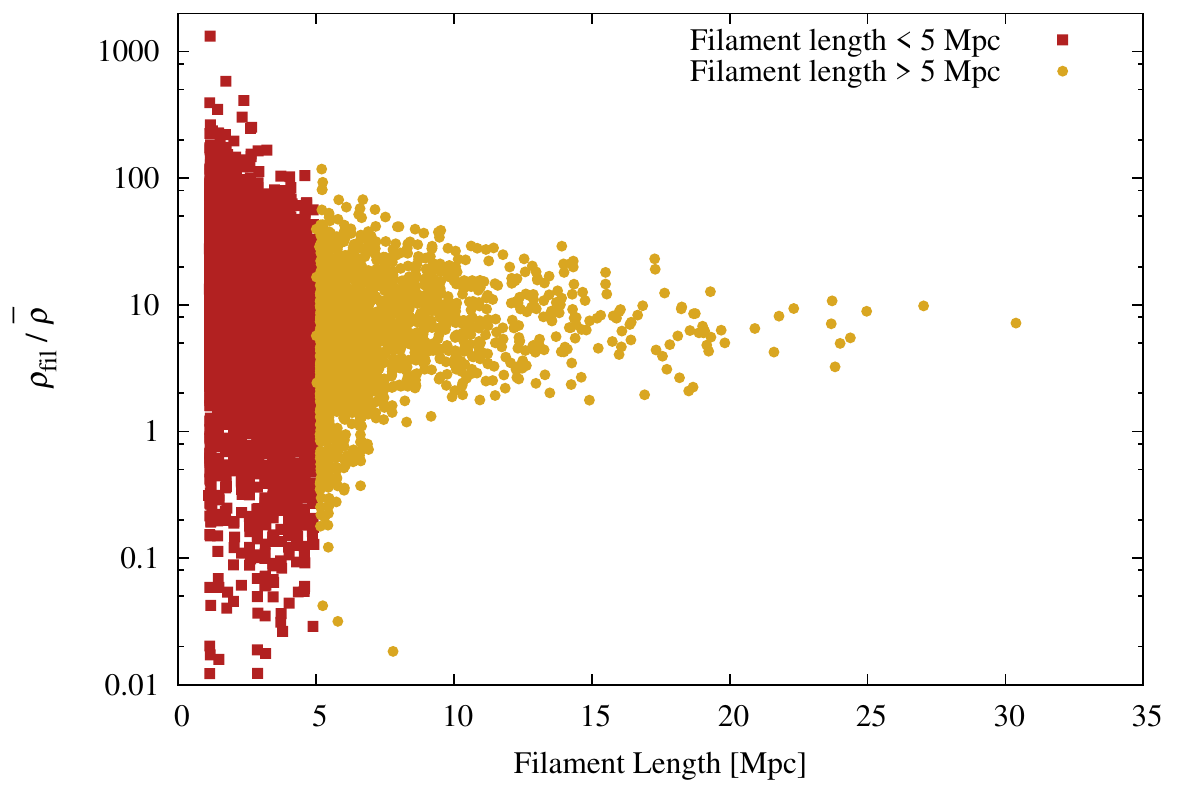}
 \caption{Filament length vs the median of the luminosity density field value with 1.5 Mpc smoothing along the filament spine. Filaments shorter than 5 Mpc (red squares) occur in a very wide range of densities and are excluded from the subsequent analysis.}
 \label{fig:flentolden}
\end{figure}

%________________________________________________________________
\section{Data}\label{sect:data}

Our dataset is based on DR12 \citep{Alam:15}, which is part of the SDSS-III project, a continuation of previous SDSS observing programmes and described in detail in \citet{Eisenstein:11}. The full flux-limited catalogue of the galaxy properties we use is available in CosmoDB\footnote{URL: \url{http://cosmodb.to.ee}} and the steps used for cleaning the galaxy sample and for galaxy group determination are described in the catalogue of \citet{Tempel:17} (hereafter T17). Galaxies in the catalogue are selected from the main contiguous area of the survey which covers 7221 square degrees of the sky. Absolute luminosities of galaxies in $u, g, r, i, \mathrm{and} z$ filters are k- \& e-corrected. 

The measured properties in this study are gathered from publicly available DR12 data. We use $g - i$ and $g - r$ colours calculated from the corrected luminosities in the T17 catalogue. From the DR12 SkyServer database \citep{Alam:15} from the \textit{galSpecExtra} table, we use stellar mass defined as the median estimate using model photometry and SSFR defined as the median estimate from emission line measurements combined with model photometry. From the \textit{galSpecIndx} table we use D4000 determined using the \citet{Balogh:99} definition after correction for emission lines. The definitions are described in \citet{Brinchmann:04}. SSFR is an estimate of overall star formation activity in a galaxy. Its numeric value depends on model spectra and involves more assumptions. D4000 is measured directly from the observed spectra, but is a tracer for only the most recent star forming period of a galaxy. Galaxy morphology classification is taken from a catalogue based on the SDSS DR10. The catalogue is described in \citet{Tempel:14d} (hereafter T14). Galaxies are classified as having elliptical, spiral, or undefined morphology (described in detail in \citealt{Tempel:11}). \textit{Specobjid} column matching was used to join the T17, T14, and \citep{Alam:15} datasets. 

Groups in the T17 catalogue were detected using the friends-of-friends (FoF) method with membership refinement as described in \citet{Tempel:16b}. In brief, the FoF method, which might leave sub-systems of groups that are near each other undetected, was complemented by a multi-modality check on the detected groups. Additionally the virial radius and escape velocity of each group was calculated to separate unbound galaxies. The flux-limited sample along with a varying linking length was used when determining groups. Group properties were found to be roughly constant with redshift. 

To ensure galaxy selection completeness we use a volume-limited sub-sample of the full flux-limited survey with a luminosity cut $M_\mathrm{r}$ $\leq -19$ mag. The luminosity cut was chosen as a balance between the final volume and a broad range of luminosities of selected galaxies. The volume-limited galaxies are selected by their group distance within $\mathrm{85 - 215 ~ Mpc}$ (redshift $\mathrm{0.02 - 0.05}$). Nearby galaxies are removed because in the local universe the uncertainties of galaxy distances and the large angular size makes the Bisous method of filament detection (described below) uncertain. The farther distance cut corresponds to the luminosity limit. In stellar mass the galaxy sample is complete down to $\log{\left(\tfrac{M_{\star}}{M_\odot}\right)} = 9.8$. The volume-limited catalogue contains 55973 galaxies. 

After the luminosity density field calculation and filament detection (described in Sect. \ref{sect:meth}) we additively removed some galaxies from further study based on their measured properties. We left out galaxies without stellar mass and SSFR estimates in the DR12 \textit{galSpecExtra} tables removing about 7\,\% of galaxies. We made a cut based on two sets of colours, $-0.5~\mathrm{mag} < g - i < 2.0~\mathrm{mag}$ and $-1.0~\mathrm{mag} < u - z < 5.4~\mathrm{mag}$, in order to remove extended wings in the colour distributions. These wings are caused by errors in luminosity measurements in one or more of the five SDSS filters. The $r$-band measurements are already cleaned of these errors in the T17 catalogue. The colour cuts remove less than 1\,\% of galaxies from the full table.

\begin{figure*}
 \centering
 \includegraphics[width=180mm]{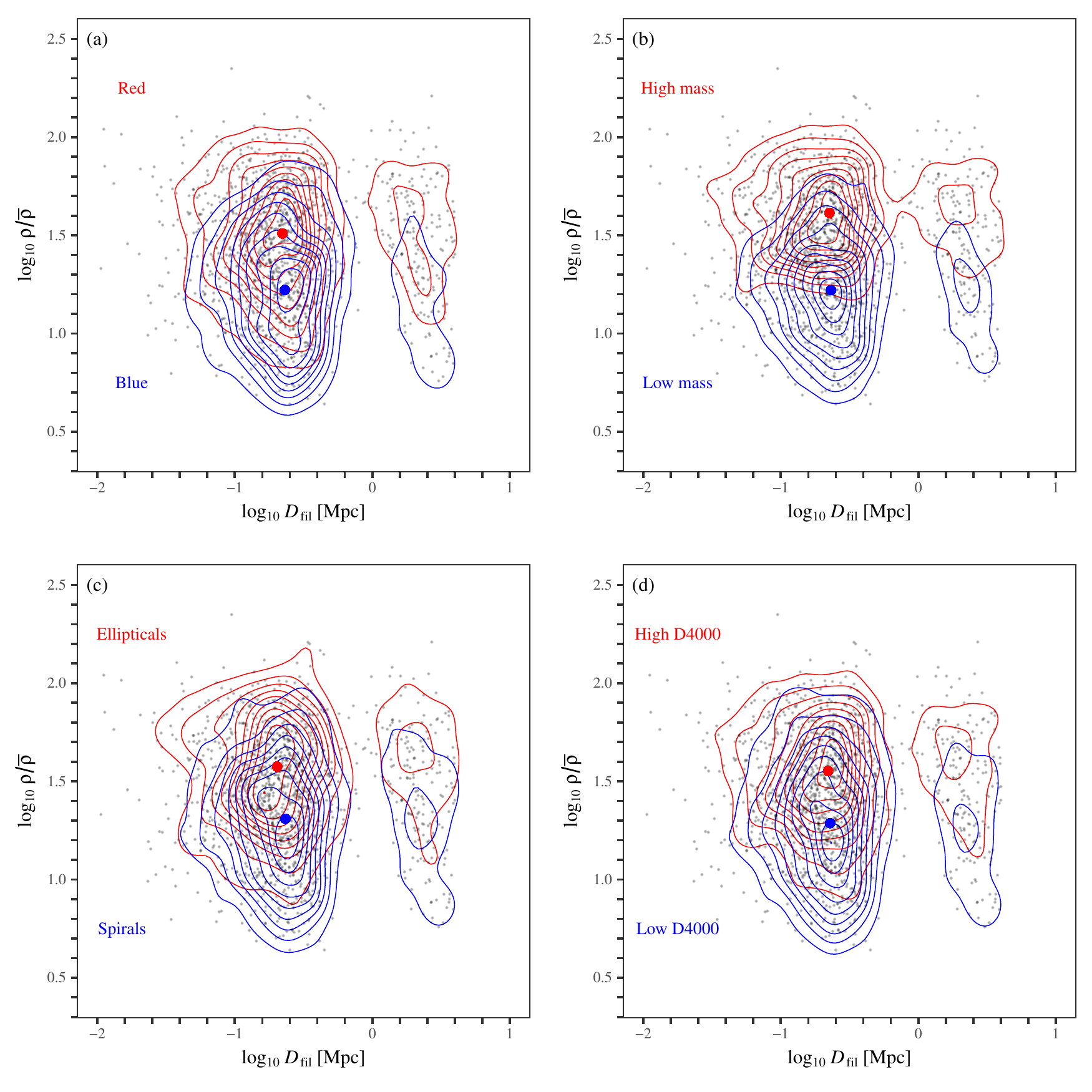}
 \caption{Relationship between BGGs $D_{\mathrm{fil}}$ and luminosity density field values for (a) red and blue, (b) high and low mass, (c) elliptical and spiral, and (d) high and low D4000 BGGs. Red and blue contours show the two-dimensional density of the separated populations. Red and blue dots show the peaks of BGG distributions in their respective populations. Individual BGGs are shown with grey points. The variation in populations mostly depends on the luminosity density field at BGG locations.}
 \label{fig:sdss_dfdencont}
\end{figure*}

\begin{figure*}
 \centering
 \includegraphics[width=180mm]{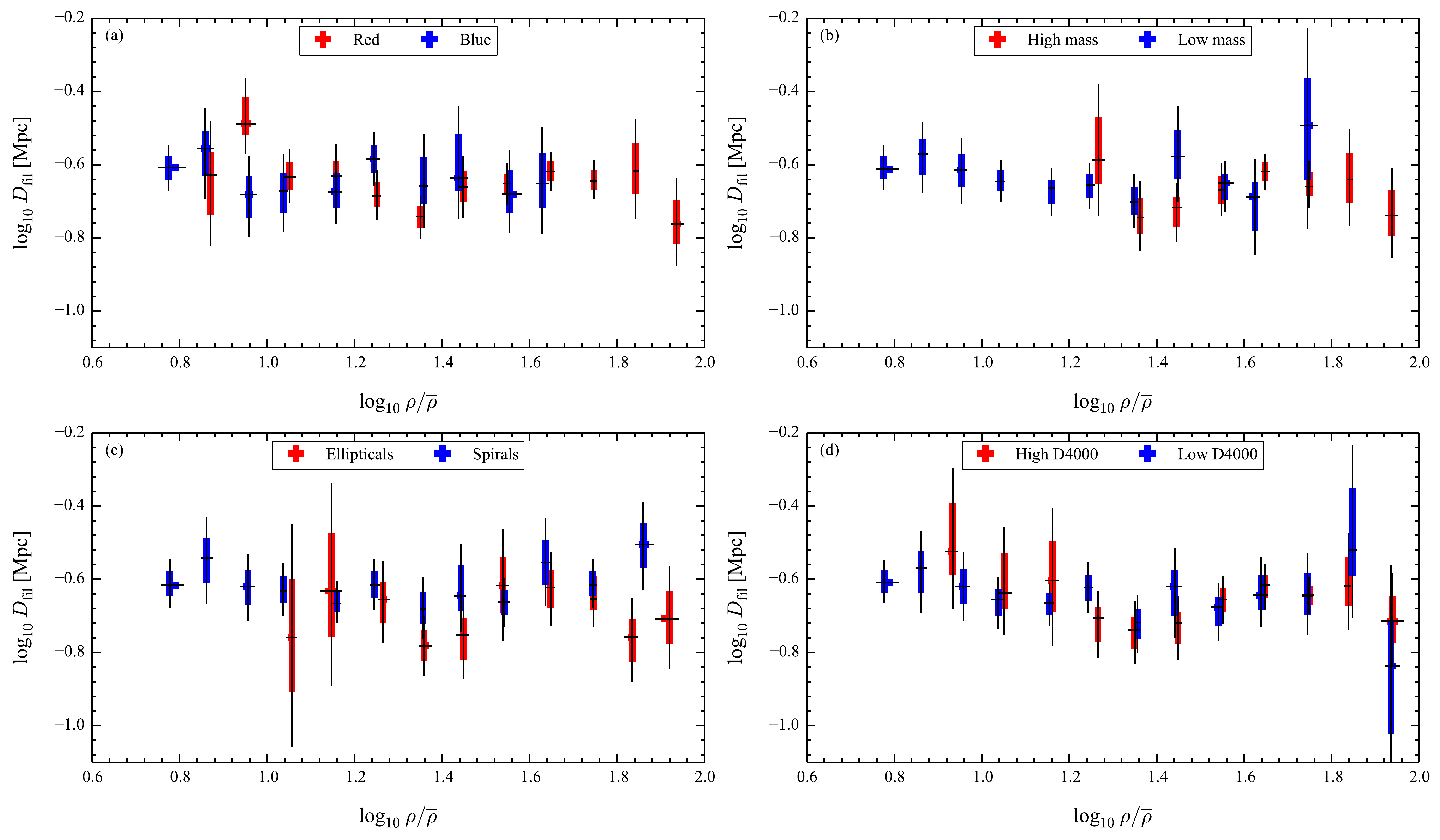}
 \caption{Median distance from the nearest filament spine of BGGs selected in density bins of constant width $\log{\left(\tfrac{\rho}{\overline{\rho}}\right)} = 0.1$. The panels show (a) red and blue, (b) high and low mass, (c) elliptical and spiral, and (d) high and low D4000 BGGs. Only bins with at least ten BGGs are shown. The thicker coloured error bars are $65\,\%$ and the thinner black bars are $95\,\%$ confidence intervals calculated using bootstrap resampling. Differences in filament distances are minor and none of the populations are consistently closer to or farther from filament spines.}
 \label{fig:sdss_bindf}
\end{figure*}

\section{Methods}\label{sect:meth}

\subsection{Filament sample}\label{sect:fila}

Filament extraction is done using the Bisous method, a marked point process designed to model multi-dimensional patterns \citep{Tempel:14a, Tempel:16a}. The basis for filament spine extraction is the three-dimensional spatial distribution of galaxies, making the Bisous method applicable to observational data. The Bisous method has shown good agreement when compared to the velocity shear web method in simulations \citep{Libeskind:15}. 

The basic methodology is the following. 
Filamentary structures are probed with cylindrical shape segments, which are fitted to and adjusted on the galaxy distribution such that the galaxies inside the cylinder radius outnumber the galaxies within two cylinder radii. Nearby cylinders that are well aligned have an attraction parameter connecting individual segments into longer chains.
The process is stochastic, so variations between different Markov chain Monte Carlo runs of the fitting creates a likelihood map of the most probable filament locations (called the visit map). By default 1000 runs are made in one instance. Spine locations are defined based on the locations of density ridges in the visit map.
The calculated spine locations we use in this study represent the most likely locations of filament spines. 

The cylindrical elements have a varying radius between 0.5 and 1 Mpc, with a distribution average at 0.75 Mpc. The filament finder is used on the volume-limited SDSS sample. The finger-of-god effect is already suppressed in the galaxy group catalogue. We tested for the Kaiser effect by comparing results based on line-of-sight Bisous filaments to filaments along the sky plane, and found no significant differences. 

In the Bisous code the perpendicular distance of a galaxy from the nearest filament spine ($D_{\mathrm{fil}}$) is calculated. We remove all galaxies to which the closest filament point is an endpoint of any filament. These galaxies are typically located either in knots or voids far from any filamentary structures and might bias the effects we are looking for in this study. This removes roughly half of all the galaxies from the selection.

\begin{figure}
 \centering
 \includegraphics[width=88mm]{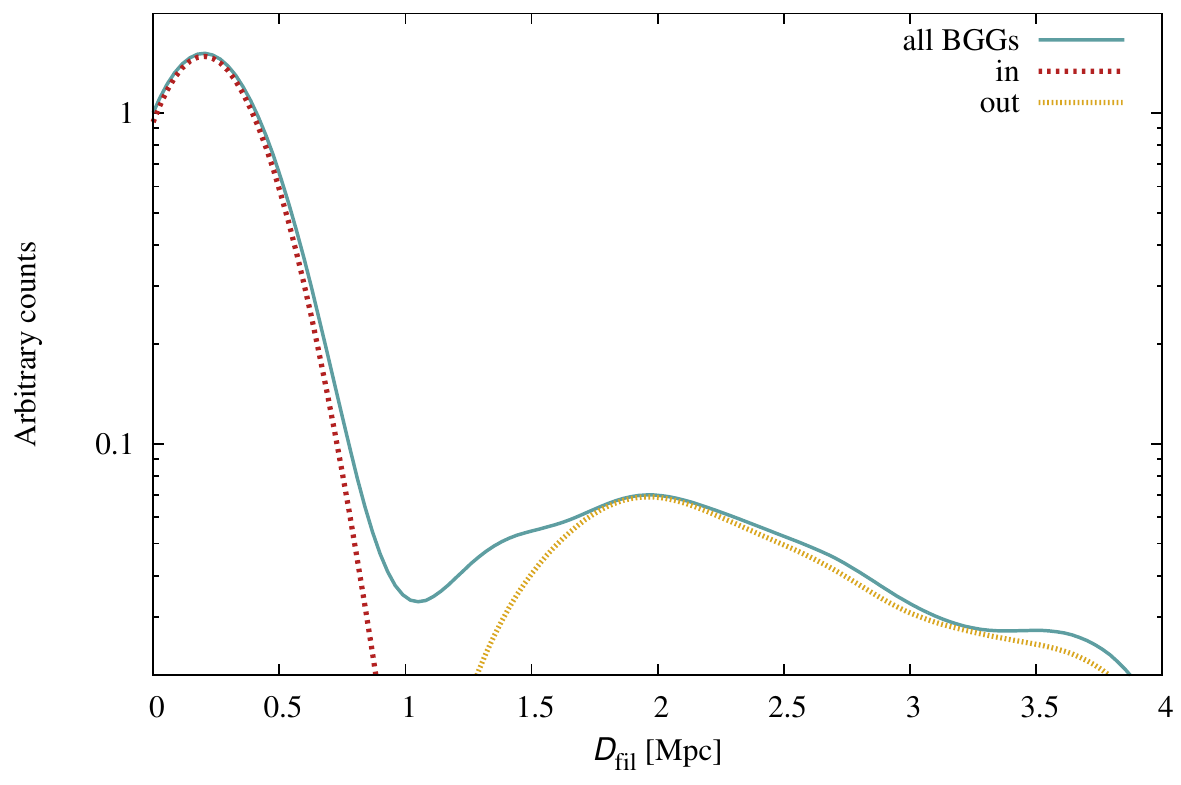}
 \caption{Gaussian-smoothed distributions of BGGs as a function of $D_{\mathrm{fil}}$. The `in' and `out' samples are defined to have no overlap.}
 \label{fig:fdist}
\end{figure}
 
\subsection{Luminosity density as environment tracer}\label{sect:lumden}

The main determining factor of galaxy properties is their local environment, measurable for example by the halo mass of the groups of a given galaxy. Care needs to be taken to neutralise this contributing factor when looking for filament-based properties. Dynamical group mass estimates available in the T17 catalogue are inaccurate for groups with only a few members. Instead, we use the luminosity density field value at the location of a given galaxy as a proxy for the mass of groups. The advantage of the luminosity density method is that it combines the number density and luminosity of the galaxies, which both relate to the underlying dark and baryonic matter mass-density field \citep[e.g.][]{Mo:04, Nevalainen:15}. Thus, luminosity density can be used as a tracer of the mass distribution in multiple scales which makes it applicable to study unvirialised structures. The luminosity density field was calculated using the $r$-band luminosities of galaxies in the volume-limited sample with a $B_{3}$ spline kernel \citep[described in detail in Appendix A in][]{Liivamagi:12}. The choice of the kernel size depends on the scale of the environment to be probed. Previous studies have used the luminosity density field with a $8~h^{-1}~\mathrm{Mpc}$ (or roughly 12 Mpc in Planck cosmology used in this paper) smoothing radius as a metric for defining the large-scale environment of galaxies \citep[e.g.][]{Lietzen:12,Poudel:16}. We use a smoothing radius of 1.5 Mpc. Previous work done using the same luminosity tracers have defined the small-scale radius as 1.5 Mpc (or the roughly equivalent $1~h^{-1}~\mathrm{Mpc}$) as an approximation of group and cluster spatial sizes \citep{EinastoJ:18}. Values of luminosity density are given as density over the mean luminosity density of the entire sample $\left(\tfrac{\rho}{\overline{\rho}}\right)$, where~$\overline{\rho}=1.11\times10^8~L_\odot~\mathrm{Mpc^{-3}}$ in the volume-limited sample. Only galaxies in the volume-limited selection ($M_\mathrm{r} \leq -19$ mag) are included in the luminosity density field calculation. 

We make a cut based on the survey edge. The luminosity density is slightly underestimated near the edges of the survey because of unobserved galaxies beyond the edge of the SDSS footprint and near and far cuts of the volume-limited sample. Making a survey edge cut corresponding to the radius of smoothing used in the luminosity density field calculation, 1.5 Mpc, removes about 5\,\% of all galaxies and mostly affects isolated galaxies not belonging to any group farther away from filaments. This is a minor effect and not significantly affecting the statistics of our results. 

\begin{figure*}
 \centering
 \includegraphics[width=180mm]{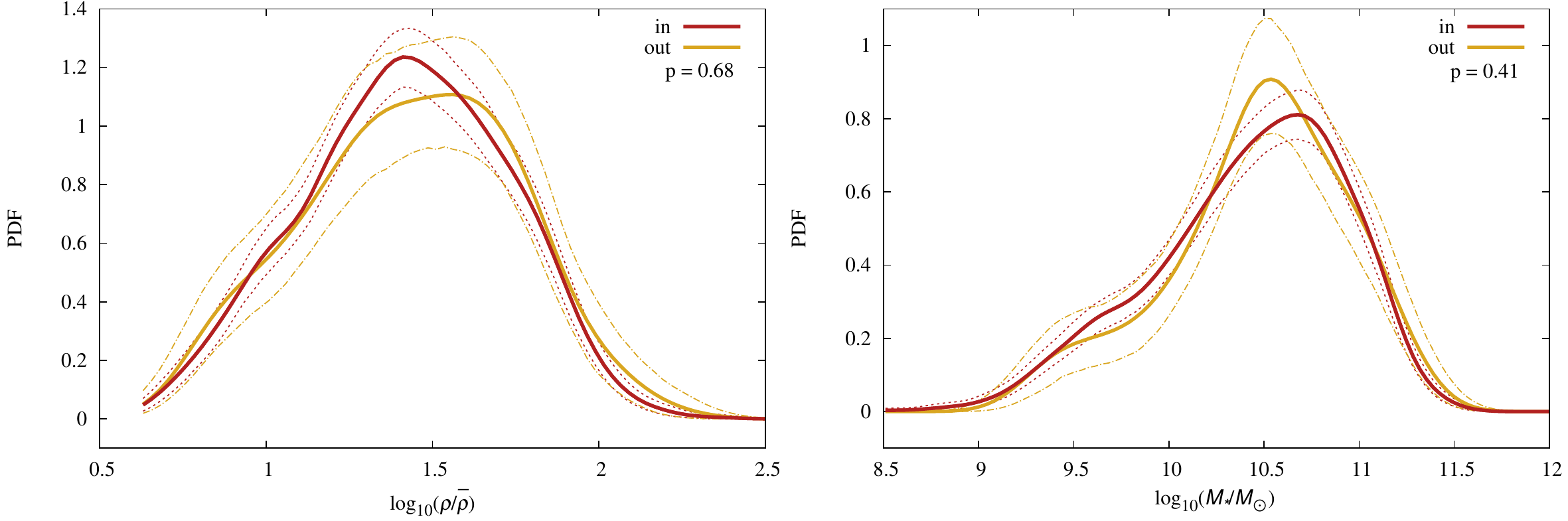}
 \caption{Gaussian-smoothed distributions of BGGs luminosity density field values (left) and stellar masses (right) in two distance bins from the nearest filament spine. The error corridors in each bin show $95\,\%$ confidence levels calculated using bootstrap resampling. A two-sample KS-test was run on the distributions in each panel. The resulting p-value of the test is shown in the plots. The `in' and `out' samples have comparable distributions of luminosity density and stellar mass.}
 \label{fig:dr12_den_mass}
\end{figure*}

\begin{figure*}
 \centering
 \includegraphics[width=180mm]{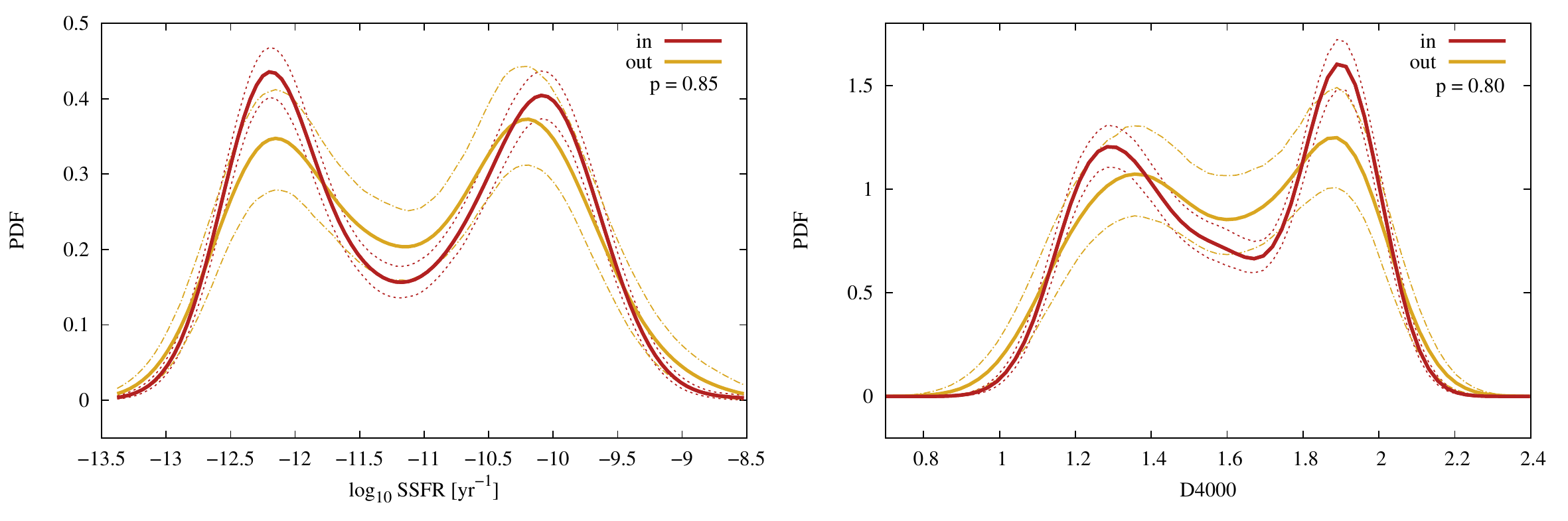}
 \caption{Gaussian-smoothed distributions of BGGs SSFR and D4000 in two distance bins from the nearest filament spine. The plots have the same notation as in Figure \ref{fig:dr12_den_mass}. The samples have minor differences, but not significantly outside the error corridors and statistically could be drawn from the same dataset.}
 \label{fig:dr12_ssfr}
\end{figure*}

To test how luminosity density traces group mass, we used data from the EAGLE simulation \citep{Crain:15,Schaye:15}. The EAGLE simulation uses a modified version of the Gadget-3 Smoothed Particle Hydrodynamics code \citep{Springel:05b} and a series of sub-grid models for implementing physical processes occurring on small unresolved scales. The dark matter halos are initially identified from the dark matter particle distribution using the FoF algorithm and a linking length of 0.2 times the average inter-particle distance. Galaxies are then identified from the baryonic matter particle distribution within halos using the SUBFIND algorithm \citep{Springel:01,Dolag:09}. We used the simulation run Ref-L100N1504 in a cubic box 100 Mpc in width, which is large enough to characterise the large-scale environment of galaxies. We use $r$-band magnitudes of galaxies to determine the brightest galaxy of each FoF halo and to calculate the luminosity density field. For each galaxy we used magnitude with dust attenuation (described in \citealt{Camps:18}) if dust attenuated magnitudes were available, otherwise dust-free magnitudes were used. We checked distributions of combinations of colours for galaxies with and without dust attenuation and found a closer match to the SDSS sample colours when using dusty magnitudes in EAGLE. Galaxy positions in the simulation box were based on their centre of mass. For group mass we used the total mass of each FoF halo. We selected galaxies using the same average number density as for the SDSS volume limited catalogue ($8.187\times10^{-3}~\mathrm{Mpc}^{-3}$), corresponding to a luminosity cut of $M_\mathrm{r} \leq -18.4$ mag. This was done to ensure the same luminosity density normalisation for both samples. The luminosity density field for EAGLE galaxies was calculated in the same way as for the observed galaxies. The periodic box allowed us to ignore survey edge effects that had to be accounted for in the observations.

We selected BGGs inside EAGLE groups with the same limitations to filament parameters that we use for observed BGGs (494 BGGs in EAGLE; the exact cuts are described below). The luminosity density distribution of these BGGs was similar to that of the observed BGGs. In Figure \ref{fig:eag_masslum} the relation between group mass and luminosity density in EAGLE is shown. There is a clear correlation between the two parameters. We note that the $1\sigma$ standard deviation is up to 0.5 dex; however, the total relation can be approximated by a linear fit shown in the figure. There is a slight deviation from a linear correlation at the lowest and highest densities. This is likely caused by the lower number of BGGs at these densities. In the analysis below we use the luminosity density value of T17 galaxies as a tracer of their group masses. We only make use of the correlation without trying to estimate the exact group masses. For comparison, the scatter of dynamical mass estimates for such poor groups in SDSS is 2-3 dex larger than those shown in Figure \ref{fig:eag_masslum}. They are also larger for richer and more massive systems. Thus, luminosity density provides a more reliable tracer of group mass.

We note that for isolated galaxies the luminosity density versus halo mass relation is not as reliable (not presented here). In the case of isolated galaxies with halo mass below $12.5\times\mathrm10^{10}~{M}_{\odot}$ the dispersion becomes wide. Thus, we cannot reliably constrain halo masses of isolated galaxies based on luminosity density field.

After selecting only BGGs from the T17 galaxies 95\,\% of them are within 4 Mpc of the nearest filament spine. Up to that distance the distribution of luminosity density field value at BGG locations is roughly constant as a function of $D_{\mathrm{fil}}$. Farther away from the spines there is a marked decrease in the median luminosity density field value. Galaxies at farther distances are mostly isolated. This shift in luminosity density distribution implies there is a cut-off distance between galaxies near larger structures and voids. We want to avoid putting these different environments together when analysing the properties of BGGs. Thus we limit our analysis to BGGs within 4 Mpc of the nearest filament spine.

By visual inspection we found that spines shorter than 5 Mpc in length tend to appear inside larger clusters or conversely as isolated segments far from other structures. We want to ignore dense cluster regions to minimise the effect that clusters have on galaxies. The sparse number of galaxies in low-density regions makes the detected shorter filaments uncertain as well. Instead, we focus on longer filaments, which are also found to be in intermediate- or low-density regions as shown in Figure \ref{fig:flentolden}. We removed galaxies for which the nearest filament is shorter than 5 Mpc from the analysis. We note that filaments in Fig. \ref{fig:flentolden} represent all filaments in the volume-limited SDSS sample. After BGG selection and group richness cuts the remaining filaments in the analysis are located at filament densities between $1 < (\rho_{fil}/<\rho>) < 100$.

In the full volume-limited sample 55\,\% of groups are pairs and 93\,\% have six or fewer members. After the selections described above these values become 62\,\% and 97\,\%, respectively. We exclude the 3\,\% of BGGs that belong to groups with more than six members, as argumented further in the results (Sect. \ref{sect:res}), yielding for the final sample 64\,\% of pairs. In general, groups in filaments have few members and we focus on those poor groups. This emphasises the need to prefer luminosity density over dynamical estimates in order to determine the local environment of BGGs in our data.
In the final SDSS sample we have 1382 BGGs in 976 filaments.

\begin{figure}
 \centering
 \includegraphics[width=88mm]{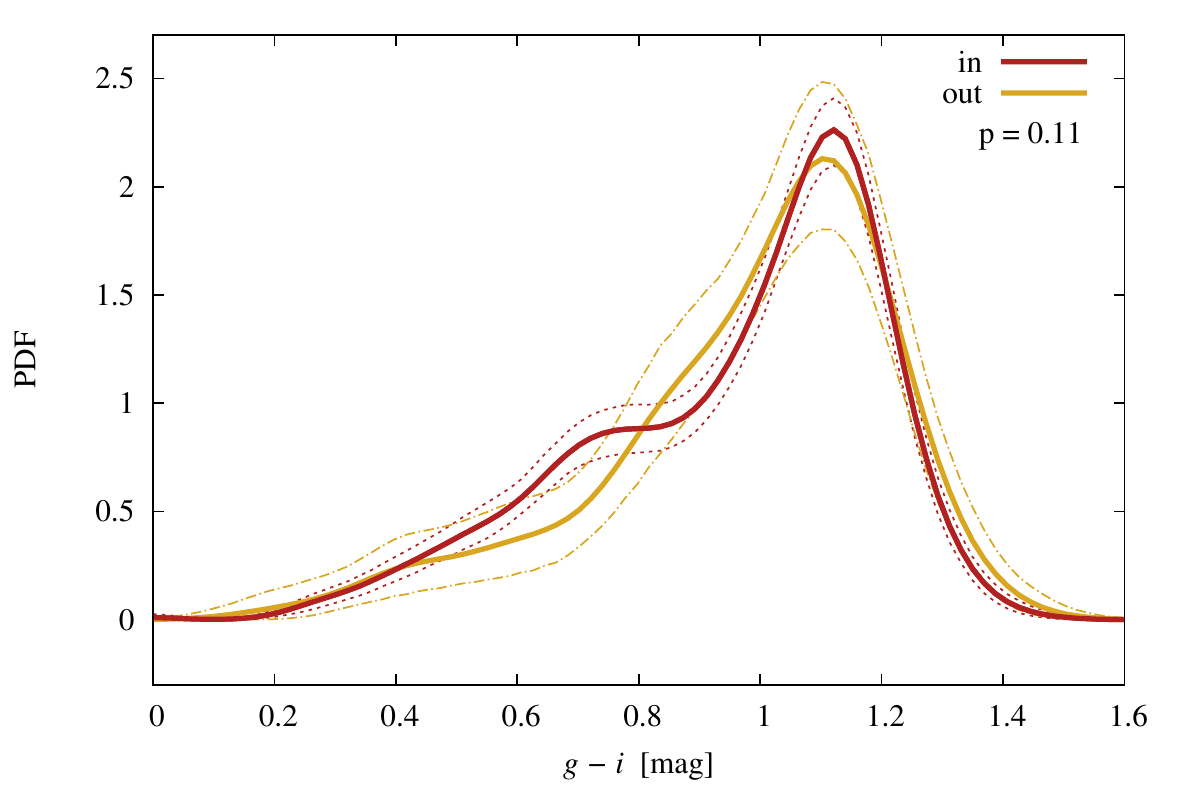}
 \caption{Gaussian-smoothed distributions of $g - i$ colours in two distance bins from the nearest filament spine. The plots have the same notation as in Figure \ref{fig:dr12_den_mass}. The samples have minor differences, but not significantly outside the error corridors and statistically could be drawn from the same dataset.}
 \label{fig:dr12_colour}
\end{figure}

\section{Results}\label{sect:res}

First we check whether galaxies with different properties tend to be closer or more distant from filament spines. 
Figure \ref{fig:sdss_dfdencont} shows the number contours of different BGG populations on the plane of $D_{\mathrm{fil}}$ and luminosity density field values. The BGGs are divided into populations based on $g - i$ colour in panel (a), stellar mass (b), morphology (c), and D4000 value (d). We use a limiting value $g - i = 0.82$ mag corresponding to the `green valley' of the $g - i$ distribution to delineate red and blue BGGs. High-mass and low-mass BGGs are separated based on the median of the stellar mass distribution at $3.2\times\mathrm10^{10}~{M}_{\odot}$. We divide BGGs into elliptical and spiral morphologies based on the classification in the T14 catalogue. To separate the D4000 distribution we use the minimum point of the bimodal distribution at D4000 = 1.6 to separate the populations that have high and low values.

For all the distributions most of the variation between the populations is along the luminosity density (vertical) axis, with red, high stellar mass, elliptical, and high D4000 BGGs having consistently higher density values, in agreement with the general understanding of the dependence on environment density. On average the peaks of the red, high stellar mass, elliptical, and high D4000 BGGs have local densities that are about two times higher than the respective blue, low stellar mass, spiral, and low D4000 distribution peaks. There is a slight variation in the peaks based on $D_{\mathrm{fil}}$, most visibly for the morphological separation in panel (c). However, the average of the elliptical population compared to the spiral population is closer to filament spines only by 0.03 Mpc. We also divided BGGs by their SSFR (not shown in the plots). The result was similar to the D4000 distribution shown in panel (d). It is clear that the environment traced by the luminosity density field value has a significantly larger effect on BGG properties compared to $D_{\mathrm{fil}}$.

In the panels of Figure \ref{fig:sdss_dfdencont} we can see a minimum in galaxy numbers at about 1 Mpc from the spine. The Bisous method places the filament cylinders so that statistically the number of galaxies between 1 and 2 cylinder radii is lower than inside 1 cylinder radius, naturally leading to the minimum in galaxy distribution as a function of $D_{\mathrm{fil}}$. The minimum occurring at one Mpc possibly characterises the size of groups in filaments, as mentioned in Section \ref{sect:lumden}, which means that there are two sets of contours per population along the $D_{\mathrm{fil}}$ axis.

Next we check how much the median $D_{\mathrm{fil}}$ of BGGs varies at fixed local densities. In Figure \ref{fig:sdss_bindf} we show the median $D_{\mathrm{fil}}$ of the same populations described in Figure \ref{fig:sdss_dfdencont}. To isolate filament distance from local environment we divide BGGs into narrow luminosity density bins of constant width $\log{\left(\tfrac{\rho}{\overline{\rho}}\right)} = 0.1$. Only bins with at least ten galaxies are shown. Errors were calculated using bootstrap resampling. The coloured bars and black lines respectively show $1\sigma$ and $2\sigma$ errors. At fixed density, there is significant $2\sigma$ overlap in median $D_{\mathrm{fil}}$ of the separated populations. There is no preference for one population of BGGs to be significantly closer to filament spines. In summary, we find no clear systematic trends between different BGG populations and $D_{\mathrm{fil}}$ at fixed environment density.

To check the effect group richness has on the differences between density-filament distance contours of various galaxy populations, we divided our sample into three sub-samples based on group richness: pairs, triplets, and more than three members (not plotted). We found that richer groups are in higher density environments and are more red, massive, elliptical, and have a higher D4000 value. However, we do not find any systematic differences between median filament distances of the separated populations of BGGs.

\begin{figure*}
 \centering
 \includegraphics[width=180mm]{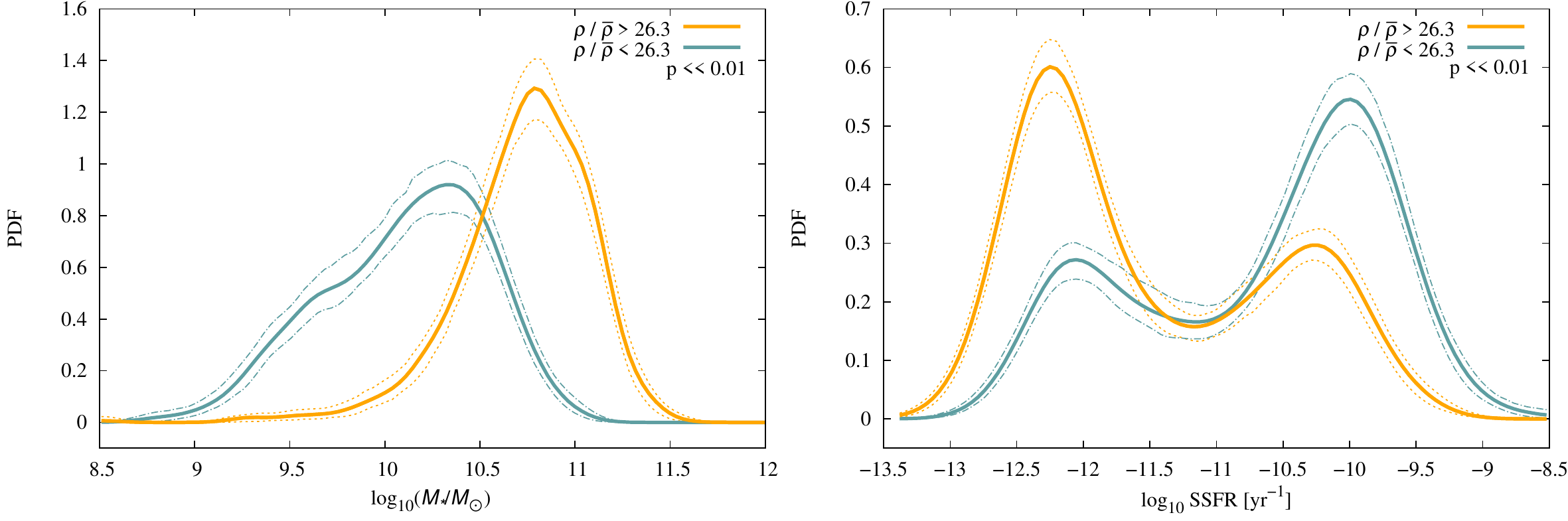}
 \caption{Distributions of stellar mass (left) and SSFR (right) of BGGs split into local high-density and low-density values. The BGGs in our analysis were split based on the mean value of the sample, with the orange line denoting high-density distributions and the blue line low-density distributions. The p-value of a KS-test on each pair of distributions is shown below the legend of each panel. BGG properties are strongly connected to their local densities.}
 \label{fig:dr12_denbins}
\end{figure*}

Figures \ref{fig:sdss_dfdencont} and \ref{fig:sdss_bindf} do not reveal significant differences of $D_{\mathrm{fil}}$ distances of different BGG populations. We can also look at it the other way round: we check whether the distributions of BGG properties change based on their $D_{\mathrm{fil}}$ value. The distribution of BGG distance from filament spines is shown in Figure \ref{fig:fdist}. The BGG sample is separated into `in' and `out' sub-samples. The former is defined to have $D_{\mathrm{fil}}$ values less than the filament radius, the latter $D_{\mathrm{fil}}$ between two times the filament radius value and 4 Mpc. A gap is left between the selections to avoid overlap between populations of BGGs at around the filament radius. The $D_{\mathrm{fil}}$ interval that we ignore has few BGGs and consequently lower confidence of measurements. The group richness distributions for the `in' and `out' samples are similar, and even more so after the exclusion of BGGs of groups with more than six members. In the `in' and `out' samples we have 1186 and 167 BGGs, respectively; limiting by richness leaves 1158 and 153 BGGs, respectively. Below we compare distributions of SSFR, D4000, $g - i$, and $g - r$ colours in these samples. 

The Gaussian-smoothed distributions of luminosity density and stellar mass are shown in Figure \ref{fig:dr12_den_mass}. Error corridors of the distributions are $95\,\%$ confidence intervals calculated from 1000 bootstrap iterations with replacement. After the sample refinement by filament properties and group richness described above, the density and stellar mass distributions are similar within the error corridors. This shows that we have comparable samples of BGGs with similar group environment and masses independent of $D_{\mathrm{fil}}$. Figures \ref{fig:dr12_ssfr} and \ref{fig:dr12_colour} show the SSFR, D4000, and $g - r$ distributions of the same BGG samples. For SSFR, shown in the left panel of Figure \ref{fig:dr12_ssfr}, the relative fraction of passive BGGs is slightly higher for the `in' sample. A similar effect is seen in the D4000 distribution. The second peak at higher D4000 values corresponds to galaxies that have a longer time passed since the last star formation activity. The $g - i$ colour distributions are shown in Figure \ref{fig:dr12_colour}. Interestingly, inside the filament radius there is a larger population of blue BGGs. For $g - r$ colours (not shown) the distributions are similar to the $g - i$ colour distributions. Overall, the distributions of galaxy properties do not vary significantly beyond the $95\,\%$ confidence corridors. 

To quantify the similarity of the two distributions a two-sample Kolmogorov--Smirnov (KS) test was run on the two samples for each property. The result of a KS test, the p-value, is a measure of the likelihood that two samples are drawn from the same distribution. Typically a p-value of less than 0.05 indicates that the samples being tested are unlikely to be drawn from the same distribution. The p-values of the test are shown in the figures. All are high enough to indicate that the samples shown in Figures \ref{fig:dr12_den_mass}-\ref{fig:dr12_colour} are drawn from the same distribution. The KS test confirms that the BGGs are similar, independent of their distance to the nearest filament spine. For comparison Figure \ref{fig:dr12_denbins} shows the effect of luminosity density on the properties of BGGs. We split the BGG sample based on the median luminosity density value $\left(\tfrac{\rho}{\overline{\rho}}\right) = 26.3$. As expected the resulting stellar mass and SSFR distributions show remarkable differences. BGGs in high-density environments are more massive and less active than BGGs in low-density environments. A KS-test on these distributions shows that they are unlikely to be drawn from the same distribution. This highlights that the local density is the main determining factor of galaxy properties.

We performed some additional tests in order to check the reliability of our results. As discussed in \citet{Kraljic:18}, the choice of the density smoothing scale can affect filament-based signals. Thus, we checked whether a larger smoothing scale, 12 Mpc ($8~h^{-1}~\mathrm{Mpc}$) would change the results presented above. With 12 Mpc smoothing the correlation between group mass and luminosity density presented in Figure \ref{fig:eag_masslum} falls apart: the group scale information is lost with oversmoothing. On the larger scale the luminosity density field traces more the supercluster-void network, the effects of which are discussed in \citet{Lietzen:12}. 

In order to assess possible selection effects we acquired flux-limited Bisous filament data for the DR12 catalogue. After limiting our selection to BGGs with the same constraints as used above and in the same distance interval, the resulting distributions in Figures \ref{fig:dr12_den_mass}-\ref{fig:dr12_colour} do not change significantly. 

Splitting the BGG sample according to group richness only reveals the relation with local density (i.e. richer groups are more massive, redder, and less active). However, the low number of richer groups increases the errors considerably, and any possible trend would be buried within the uncertainties.

From these tests we summarise that the results presented in this paper are not affected significantly by the galaxy set used to define filaments. Group richness cannot be excluded as a factor, but we postpone further analysis until more data becomes available.

%________________________________________________________________
\section{Discussion}\label{sect:disc}

We studied filament-based effects in BGGs with the aim of neutralising the effects of local environment. We used SDSS data combined with the Bisous method to identify filament spine locations. Our focus was on BGGs and filaments in intermediate-density environments in order to reduce cluster effects and misidentified filaments in low-density regions. On the one hand, our strict selection criteria for defining the galaxy sample as well as the filament sample considerably reduced the original sample size, and thus statistics. On the other hand, as shown in Figure~\ref{fig:eag_masslum} and as discussed in Section~\ref{sect:lumden}, by limiting our galaxy sample to BGGs we can use the luminosity density field smoothed over 1.5 Mpc radius as a reasonable proxy for the local environment. The subsequent analysis showed that inside and outside the filaments, BGGs live in similar local environments, making the inner and outer BGG samples directly comparable. The situation would be much different for isolated galaxies, for which the luminosity density field is not a reliable trace of local environment. We also found that after the filament sample refinement (i.e. after the exclusion of filaments shorter than 5 Mpc), our BGGs inhabit only poor groups with up to six members. 
 
In general, since the Bisous filaments are determined statistically, they are rather robust against variation. The imposed restrictions on filament length further increase the reliability of the filaments. The uncertainties of $D_{\mathrm{fil}}$ are here probed with bootstrap resampling, characterising possible variation arising from individual galaxy detections. A paper studying the uncertainty of Bisous filament locations is in preparation (Muru et al. in prep.). As indirect proof of the reliability of the Biosus method another paper studying the effectiveness of the method in tracing intergalactic gas in the EAGLE simulation is also in preparation (Nevalainen et al. in prep.).

The results show that BGGs separated by colour, mass, morphology, and D4000 are distributed similarly related to $D_{\mathrm{fil}}$. While small differences between the inner and outer population of BGGs are present the statistical significance is low, and at the $95\,\%$ confidence level the distribution functions of various galaxy properties are the same. A different result was obtained in \citet{Kuutma:17}, which looked at galaxy properties in SDSS DR10 in the context of Bisous filaments and using luminosity density as a local environment tracer of galaxies. Instead of BGGs, the whole galaxy sample was considered. It was found that bright galaxies with an absolute magnitude limit $M_r~-~5\mathrm{log}(h)~<~-20$ mag ($M_r~<~-20.8$ mag in Planck cosmology) are redder and have a larger fraction of elliptical morphologies towards filament spines. A fainter galaxy sub-sample with limits $-18~\mathrm{mag}~>~M_r~-~5\mathrm{log}(h)~>~-20$ mag ($-18.8~\mathrm{mag}~>~M_r~>~-20.8$ mag) did not exhibit this trend above the uncertainty level. We speculate that the differences compared to the present study might arise from the local environment of satellite galaxies and isolated galaxies. While in \citet{Kuutma:17} care was taken to render the environment densities similar by weighting the galaxy samples on the basis of the luminosity density field, which does not characterise the possible impact of nearest neighbours and is inadequate for (apparently) isolated galaxies. We note that the flux-limited filament sample used in \citet{Kuutma:17} was based on an older galaxy catalogue based on SDSS DR10 data, which has since been superseded by the T17 catalogue that has a more refined galaxy group determination method. In addition, the Bisous algorithm has had some refinements (notably the implementation of a variable radius for the cylinders forming the filamentary structure). These aspects must be the main reasons for the differences between the results of this paper and T17.

\citet{Poudel:17} looked at BGG properties related to filament distance and found larger differences than exposed here. At fixed luminosity densities, filament BGGs were found to have increased masses and decreased SSFR activity compared to non-filament BGGs. Again, the discrepancy in the results might be caused by the differences in the methodology. In the latter work, BGGs of groups with at least five members were considered (to ensure a reliable group mass estimation based on group dynamics). Looking at the current study as an extension towards smaller mass groups it appears that the trends detected in the previous study disappear. At the same time, a larger smoothing radius ($8~h^{-1}~\mathrm{Mpc}$) was applied when calculating the luminosity density field. At these scales the luminosity density field is sensitive to differences between supercluster and void environments, but not to the group and cluster scale. In addition, the sample included BGGs close to filament endpoints, which are often dominated by richer groups and clusters, and where the precise location and the effects of filaments are hard to define. In contrast, cluster environments and their proximity were intentionally avoided in the present study to have a better understanding and control over the local environment, using poor group BGGs up to a distance of four Mpc from the filament spines and filaments longer than five Mpc. 

It appears that in \citet{Kuutma:17} and in \citet{Poudel:17}, the differences in galaxy properties caused by the filament environment cannot be conclusively separated from the effects of the local environment or nearby clusters. It is also important to note that in these studies the Bisous filament finder was applied to the flux-limited SDSS dataset. As shown in \citet{Kuutma:17}, this filament sample induces undesirable redshift dependences resulting from the decreasing number density of galaxies with increasing distance; an attempt was made to reduce these effects by weighting. In the present paper we took a more conservative approach, using filaments based on a volume-limited galaxy sample.

We checked whether our results vary when changing the smoothing scale, considering filaments determined using a flux-limited galaxy data, and when separating our sample based on group richness. These tests were aimed at comparing results between this and previous works using data selection more similar to those in the previous papers. The results did not vary significantly based on these tests, or were inconclusive because of larger errors in the case of group richness separation. This shows that our findings about poor group BGGs are indeed properties intrinsic to these types of galaxies.

Several other studies have looked for changes in the properties of galaxies in relation to filaments (as overviewed in the introduction of this paper). These studies have covered a variety of surveys, redshifts, and types of galaxies. Most of the works conducted in this area find filaments to be a separate environment having some measurable effect on the properties of galaxies beyond those of the local environment. Qualitatively, galaxies in filaments appear to be more massive and less active than their counterparts outside. Because of differences in filament determination methods and in the definition of local environment, a direct comparison of results is a complicated task. It is likely that much of the variation in the results can be attributed to differences in filament definition. No universal prescription exists for defining and detecting filaments, also illustrated by the comparison of various cosmic web classification techniques in \citet{Libeskind:18} and \citet{Rost:20}. Likewise, the definition of local environment and ways of reducing its interfering impact can vary significantly between different studies. 

The results of the present study agree with studies that attribute change in galaxy properties solely to the effects of local environment \citep[e.g.][]{Yan:13, Eardley:15, Brouwer:16}. While tidal forces and gas infall from preferred directions can have a small but statistically significant effect on galaxy alignments \citep[e.g.][]{Tempel:13b}, these effects do not leave a significant impact on galaxy groups hosted by the filaments, at least not on their brightest galaxies.

In this work we focused on BGGs of groups of galaxies. The luminosity density field used in this paper is not suitable for defining the local environment of group satellites and isolated galaxies. For satellites the 1.5 Mpc radius smoothed luminosity density field may underestimate the local environment, especially at the outskirts of larger groups. Fainter satellites are just not visible at greater distances in observational surveys. Thus, isolated galaxies in observations cannot be assumed to have a fixed environment. A more detailed selection of truly isolated galaxies (i.e. using a constrained local volume sample of galaxies) would be needed in order to study these populations. We leave the study of satellites and isolated galaxies for future work.

In summary, our study has led us to the following conclusions:
\begin{enumerate}
\item Poor groups are a typical environment for galaxies in and near filaments. 
\item The luminosity density field (with 1.5 Mpc smoothing radius) is a good proxy for the local environment (in other words halo mass) of poor groups.
\item The local environment of poor groups is insensitive to the distance to the filament spine. 
\item Filament environment has a very limited effect on the properties of the brightest (poor) group galaxies; overwhelmingly, the properties of these galaxies depend on the local environment density. 
\end{enumerate}

Our approach of constraining the environment of galaxies in and near filaments leads to small sample sizes. This approach shows the limit of the SDSS volume-depth balance. Data from the next generation of galaxy redshift surveys (4MOST, DESI, Euclid, J-PAS) will be helpful for a proper assessment of any possible weak filament-specific impact on galaxy properties.

%________________________________________________________________
\begin{acknowledgements}
We thank Jukka Nevalainen and Toni Tuominen for helpful discussion during writing.
This work was supported by grants IUT26-2, IUT40-2 and PUT1627 funded by the Estonian Research Council, 
by the Centre of Excellence ‘Dark side of the Universe’ (TK133), the grant MOBTP86 financed by the European Union through the European Regional Development Fund, and the Vilho, Yrj\"o and Kalle V\"ais\"al\"a Foundation of the Finnish Academy of Science and Letters.
Funding for SDSS-III has been provided by the Alfred P. Sloan Foundation, the Participating Institutions, the National Science Foundation, and the US Department of Energy Office of Science. The SDSS-III web site is http://www.sdss3.org/.
We acknowledge the Virgo Consortium for making their simulation data available. The EAGLE simulations were performed using the DiRAC-2 facility at Durham, managed by the ICC, and the PRACE facility Curie based in France at TGCC, CEA, Bruyères-le-Ch\^atel.
\end{acknowledgements}

\bibliographystyle{aa} % style aa.bst
\bibliography{mybib} % your references Yourfile.bib

\end{document}